\documentclass{aa}
\usepackage{graphicx}
\usepackage{txfonts}

\usepackage{graphics}

\begin{document}

    
\title{Observational studies of Cepheid amplitudes. I\\
Period-amplitude relationships for Galactic Cepheids and 
interrelation of amplitudes}

\author{P\'eter Klagyivik\inst{1}$^,$\inst{2} \and 
L\'aszl\'o Szabados\inst{2}}

\institute{Lor\'and E\"otv\"os University, Department~of Astronomy, 
H-1518 Budapest, P.O. Box~32, Hungary 
\and
Konkoly Observatory, H-1525 Budapest XII, P.O. Box 67, Hungary}

\offprints{P. Klagyivik, \email{klagyi@konkoly.hu}}

\date{Received  / Accepted }
                                                                                
\abstract
{
The dependence of amplitude on the pulsation period differs from other 
Cepheid-related relationships.
}
{
We attempt to revise the period-amplitude ($P$-$A$) relationship of 
Galactic Cepheids based on multi-colour photometric and radial velocity data.
Reliable $P$-$A$ graphs for Galactic Cepheids constructed for the 
$U$, $B$, $V$, $R_{\rm C}$, and $I_{\rm C}$ photometric bands and 
pulsational radial velocity variations facilitate investigations 
of previously poorly studied interrelations between observable amplitudes. 
The effects of both binarity and metallicity on the observed amplitude,
and the dichotomy between short- and long-period Cepheids can both be studied.
}
{
A homogeneous data set was created that contains basic physical and 
phenomenological properties of 369 Galactic Cepheids. 
Pulsation periods were revised and amplitudes were 
determined by the Fourier method. $P$-$A$ graphs were constructed and 
an upper envelope to the data points was determined in each graph. 
Correlations between various amplitudes and amplitude-related parameters 
were searched for, using Cepheids without known companions.
}
{
Large amplitude Cepheids with companions exhibit smaller photometric amplitudes 
on average than solitary ones, as expected, while s-Cepheids pulsate with 
an arbitrary (although small) amplitude. The ratio of the observed radial velocity 
to blue photometric amplitudes, $A_{V_{\rm RAD}}/A_B$, is not as good 
an indicator of the pulsation mode as predicted theoretically. 
This may be caused by an incorrect mode assignment to a number of small amplitude 
Cepheids, which are not necessarily first overtone pulsators.
The dependence of the pulsation amplitudes on wavelength is used 
to identify duplicity of Cepheids. More than twenty stars previously 
classified as solitary Cepheids are now suspected to have a companion.
The ratio of photometric amplitudes observed in various bands confirms 
the existence of a dichotomy among normal amplitude Cepheids. 
The limiting period separating short- and long-period Cepheids is 10.47 days.
}
{
Interdependences of pulsational amplitudes, the period dependence of the 
amplitude parameters, and the dichotomy have to be taken into account 
as constraints in modelling the structure and pulsation of Cepheids. 
Studies of the $P$-$L$ relationship must comply with the break 
at 10\fd47 instead of the currently used `convenient' value of 10 days.
}
{}

\titlerunning{Interrelations of Cepheid amplitudes}
\authorrunning{Klagyivik \& Szabados}
                                                                                
\maketitle
                  
                                                              
\section{Introduction
\label{Sect_intro}}

Cepheid variables are considered to be among the most important 
stars to both astrophysics and establishment of the cosmic distance scale. 
Their pulsation period, $P$, eigenperiod of free radial oscillation 
(or its overtone) developing in the star, depends on the average density, 
$\rho$, of the star, according to the well known formula 
$P \sqrt{\rho} = Q$, where $Q$ is practically constant for a given type of 
pulsator (neglecting the slight dependence on stellar mass). The 
existence of the period--luminosity ($P$-$L$) relationship of Cepheids
is implied by this formula and because this pulsation is maintained 
in a narrow, nearly vertical region (referred to as the instability strip) 
in the Hertzsprung-Russell diagram.

The pure radial pulsation gives rise to a number of other
relationships for Cepheids, e.g., between the period and radius, 
period and colour, and period and age. These relationships are usually 
expressed as a function of the decimal logarithm of the period. 
This representation is reasonable because in such a way 
most relationships are linear.
There exists, however, an obvious exception: the dependence of the 
{\em pulsational amplitude} on $\log P$ is neither linear, 
nor single valued, i.e., a wide range of amplitudes is possible at a given 
pulsation period. Even the range of the pulsation amplitudes 
is a complicated function of the period. The peak-to-peak amplitude 
of variations during a complete cycle of the pulsation, a characteristic 
property of a Cepheid, provides important information about 
the energy of the pulsation, and the pattern of the period-amplitude 
($P$-$A$) graph plotted for Cepheids is specific to the host galaxy.

Several features of the period dependence of 
pulsation amplitudes can be explained qualitatively by the
physical properties of Cepheids. Longer period Cepheids are
more luminous and have a lower surface gravity, therefore
they pulsate in general with a larger amplitude. However, longer
period Cepheids are of lower effective temperature. The
longer the period, the deeper the relevant hydrogen and 
helium partial ionization zones responsible for driving 
radial pulsation. At a certain depth, convection
occurs that dampens the ordered pulsational motion. As a result,
the longest period Cepheids pulsate with a smaller amplitude 
than their shorter period (say 15-30 day) counterparts.

Previous empirical studies have shown that the photometric amplitude is 
a function of the position of the Cepheid within the instability region: 
amplitudes are largest near the blue side of the strip, and become gradually 
smaller toward the red edge (Sandage et~al. \cite{Sandetal04}; Turner et~al. 
\cite{Tetal06a} and references therein). Theoretical calculations modelling 
the pulsation of Cepheids have been unable to fully reproduce the $P$-$A$ diagram 
delineated by observational data. Model calculations by Szab\'o et~al. 
(\cite{Szetal07}), however, confirm theoretically that the spread in 
photometric amplitude at a given pulsation period is partly caused 
by differences in the location of the datapoints within the instability strip
(see their Fig.~7). The peak-to-peak radial velocity amplitudes, however, 
have not been studied since Joy (\cite{J37}). 

Cepheid-related relationships are recalibrated from time to time but
the strange $P$-$A$ relationship is an exception. Ample photometric data 
obtained in the past few decades are of higher precision than samples 
used for similar studies 3-4 decades ago; they have facilitating the revision 
of the dependence of amplitude on the pulsation period and other properties,
which is the specific purpose of this paper. Here we study the effect 
of binarity on the amplitude of pulsation, while the effect of metallicity 
will be investigated in Paper~II.

A newly compiled homogeneous database containing physical and phenomenological 
properties of Galactic Cepheids is described in Sect.~\ref{sect_sample}.
The $P$-$A$ relationships based on amplitudes in 5 photometric bands and 
radial velocity data are discussed in Sect.~\ref{sect_per-amp}. 
Section~\ref{sect_disc} deals with newly introduced amplitude parameters
and a discussion of relationships between these amplitudes. 
The conclusions are drawn in Sect.~\ref{sect_concl}.

\section{Data sample
\label{sect_sample}}

\subsection{Content of the database
\label{ssect_database}}
We collected observational data of 369 Galactic classical Cepheids.
From these data, we determined amplitudes of each Cepheid 
and derived some parameters characterizing the amplitudes. 
Cepheids with varying pulsational amplitudes were excluded 
from this study. In addition to more than 20 double-mode radial pulsators, 
we omitted the star V473~Lyrae, whose pulsation exhibits a modulation 
of a cycle length as long as 1258 days (Cabanela \cite{C91}), 
and Polaris, whose extremely low and changing amplitudes 
(Turner et~al. \cite{Tetal05}) would result in too large uncertainties 
when forming amplitude ratios. Two Galactic beat Cepheids, however, 
were used for checking some of our results (see Sect.~\ref{sect_disc}).
This sample of Galactic classical Cepheids is otherwise complete 
to the limit of 10$^{\rm m}$ average brightness in $V$. 
Some Cepheids of about 10-11th magnitude in $V$ band 
were not included because of a lack of photometric data, 
although fainter Cepheids with known spectroscopic [Fe/H] values 
and/or reliable radial velocity phase curves do occur in the database. 

This data base, published in Table~1, available at the CDS, 
contains the following information:\\
Column~1: name of the Cepheid;\\
Cols.~2-3: Galactic longitude and latitude (taken from the {\em SIMBAD} 
data base);\\
Col.~4: pulsation period (in days);\\
Col.~5: mean apparent brightness in $V$ band;\\
Cols.~6-10: peak-to-peak amplitudes in $U$, $B$, $V$, $R_C$, and $I_C$ bands, 
respectively;\\
Col.~11: peak-to-peak amplitude of the radial velocity variations 
(corrected for the effect of orbital motion in the case of known 
spectroscopic binaries);\\ 
Col.~12: ratio of radial velocity to photometric $B$ amplitudes, 
$q$ (see Sect.~\ref{ssect_ar});\\
Cols.~13-14: the $m$ parameter (to be defined in Sect.~\ref{ssect_slope}) 
characteristic of the wavelength dependence of the photometric amplitude 
and its uncertainty;\\
Cols.~15-16: the $k$ parameter (to be defined in Sect.~\ref{ssect_cawd}) 
characteristic of the wavelength dependence of the photometric amplitude 
and its uncertainty;\\
Col.~17: iron abundance, [Fe/H];\\
Col.~18: binarity status: 0: no known companion, 1: known binary 
(or more than one known companion);\\
Col.~19: mode of pulsation: 0: fundamental mode, 1: first overtone.

Although our data base involves a smaller number of Cepheids 
than previous ones compiled by Fernie~et~al. (\cite{FBES95}, 
referred to as the DDO database), Szabados (\cite{Sz97}), and 
Berdnikov~et~al. (\cite{BDV00}), it is homogeneous and contains 
more information about the stars including [Fe/H] values, information about 
binarity and pulsational mode. In the case of several Cepheids, some fields 
have remained blank in Table~1, because of the large uncertainty in a 
given quantity (radial velocity amplitudes based on early data and 
photometric amplitudes in the $U$ band) derived from existing observations.

\subsection{Source and determination of the tabular data}
\label{ssect_source}

\subsubsection{Pulsation period}
\label{sssect_period}

Since the pulsation period of Cepheids is affected by changes 
partly due to stellar evolution, especially in the case of periods 
longer than 10$^{\rm d}$ (i.e. luminous, therefore rapidly evolving
Cepheids), special care was taken to use the true period  
for the epoch of photometric data from which amplitudes 
were determined (in general, these period values were effective 
in the 1990s). Periods were deduced by a Fourier-type 
periodogram analysis (see Sect.~\ref{sssect_amplitude}) and
rounded to 3 decimal figures in Table~1.

The pulsation period of more than 70 Cepheids in our sample 
differs from the value given in the GCVS (Samus et~al. \cite{Setal04}) 
to the third decimal place. Evolutionary or other secular period changes 
result in differences smaller than two thousandth parts of the period 
with respect to the catalogued value decades ago. 
However, the period listed in the GCVS deviates considerably from 
the true value for \object{CU~Ori} (1\fd864 instead of 2\fd160 
given in the GCVS), \object{V510~Mon} (7\fd457, instead of 7\fd307), 
and \object{CI~Per} (3\fd297 instead of 3\fd378). 
In these cases, periods given by the GCVS were determined from data 
covering one or two seasons. We involved all photometric data 
in the period analysis when determining the true value.

We did not convert the true period of first overtone Cepheids 
into the corresponding fundamental mode period, in contrast to 
the approach of Berdnikov~et~al. (\cite{BDV00}). The fundamental 
period that corresponds to the first overtone periodicity 
can be calculated using the conversion formula
\begin{equation}
P_1 / P_0 = -0.0143 \times \log P_0 - 0.0265 \times {\rm [Fe/H]} + 0.7101
\label{fundamentalization}
\end{equation}
\noindent as derived by Szil\'adi et~al. (\cite{Szietal07}). 

\subsubsection{Pulsation amplitudes}
\label{sssect_amplitude}

In some cases, widely differing amplitudes are listed for the same Cepheid 
in various sources. To avoid using erroneous data, we redetermined 
the amplitudes from the original observational data. 
If Berdnikov (\cite{B08}) and his coworkers had not observed 
the given Cepheid or if the unfavourable phase coverage resulted in 
an unacceptable value, other photometric series, obtained mainly 
by Coulson \& Caldwell (\cite{CC85}), Coulson et~al. (\cite{CCG85}), 
Gieren (\cite{G81}, \cite{G85}), and Moffett \& Barnes (\cite{MB84}) 
were analysed. Amplitudes taken from Szabados (\cite{Sz97}), 
also derived by Fourier decomposition, 
are listed for up to two decimal places in Table~1.

Most radial velocity amplitudes were determined from data published 
following the last update of the tables in the DDO database, the main 
sources of which were Barnes et~al. (\cite{Betal05}), Bersier (\cite{B02}), 
Bersier et~al. (\cite{Betal94}), Gorynya et~al. (\cite{Getal92}; 
\cite{Getal96}), Groenewegen (\cite{G08}), Imbert (\cite{I99}), 
Kienzle et~al. (\cite{Ketal99}), Kiss (\cite{K98}), Petterson et~al. 
(\cite{Petal05}), Pont et~al. (\cite{Petal94}), and Derekas (personal 
communication). More recent data infer a scatter of about 1-2 per cent in the 
radial velocity phase curve. Earlier radial velocity data obtained 
by Joy (\cite{J37}), Stibbs (\cite{S55}), Feast (\cite{F67}), and
Lloyd Evans (\cite{LE68}, \cite{LE80}) were taken into account 
if more recent data were not available for a given Cepheid.

The amplitudes were determined by decomposing the phase curves into 
Fourier terms using the program package MuFrAn (Koll\'ath \cite{K90}). 
In the Fourier decomposition, the observed time series was fitted
by the sum of sinusoidal terms with frequencies corresponding to the 
observed pulsation period and its harmonics. In the case of monoperiodic 
variable stars, the instantaneous brightness value can be written as
\begin{equation}
m(t) = A_0 + \sum_{i=1}^n A_{i}\cos [i\omega (t-t_0) + \phi_i] ,
\label{four_decomp}
\end{equation}
\noindent where $t$ is time counted from an arbitrary $t_0$ moment, 
and the coefficients $A_i$ and  $\phi_i$ represent the amplitude and 
the phase of the corresponding term in the Fourier expansion, respectively, 
while $\omega = 2\pi /P$, where $P$ is the observed pulsation period.

The shape of the light curve can be described quantitatively by properly 
defined parameters based on Fourier coefficients. 
The most useful set of parameters was proposed 
by Simon \& Lee (\cite{SL81}). Following their suggestion and notation, 
the amplitude ratios, $R_{ij}=A_i/A_j$, and the $\phi_{ij}=j\phi_i-i\phi_j$ 
phase differences are commonly investigated. 
In spite of more recent interest in the $R_{ij}$ and 
$\phi_{ij}$ Fourier parameters, here we study only the peak-to-peak 
amplitudes. The behaviour of the $R_{ij}$ and $\phi_{ij}$ parameters 
will be studied in a later paper.

When decomposing photometric and radial velocity phase curves, 
amplitudes were derived from the fundamental period and its 
first four harmonics but for Cepheids with a complicated light curve 
shape the fit was extended to two more harmonics.
Possible systematic differences between the amplitudes obtained 
from fits involving different numbers of harmonics were also studied. 
In the case of well covered phase curves, differences between 5- and 
7-harmonic fits turned out to be insignificant.
The goodness of the fit is very sensitive to the deviations 
from the true period. This is why special care was taken to use 
the value of the period valid for the epoch of observations analysed 
(see Sect.~\ref{sssect_period}).

We decided to use data for the $R$ and $I$ bands of the Kron-Cousins system.
The linear transformation formulae between the amplitudes in the Johnson 
and Kron-Cousins systems, based on Cepheids observed in both systems 
(93 stars in $R$ and 91 stars in $I$ bands), are as follows:
\begin{equation}
 A_{R_{\rm C}} = 1.157(\pm 0.008) \times A_{R_{\rm J}}
\label{RCRJ}
\end{equation}
\begin{equation}
 A_{I_{\rm C}} = 1.175(\pm 0.012) \times A_{I_{\rm J}}
\label{ICIJ}
\end{equation}
\noindent where $A_{R_{\rm C}}$ and $A_{I_{\rm C}}$ are amplitudes in the 
Kron-Cousins system, and $A_{R_{\rm J}}$ and $A_{I_{\rm J}}$ are
in the Johnson system. 

For Cepheids belonging to spectroscopic binary systems, 
the amplitude of the radial velocity variations of pulsational origin
was determined by removing the orbital effect from the radial velocity
data based on the orbital elements available in the online database\footnote{http://www.konkoly.hu/CEP/intro.html}
of binary Cepheids (Szabados \cite{Sz03b}).

\subsubsection{Iron abundance}
\label{sssect_ironab}

We characterize metallicity in terms of [Fe/H] values. Conventionally, 
[Fe/H] = $\log ({\rm Fe/H}) - \log ({\rm Fe/H})_{\odot}$, 
is the logarithmic {\em iron abundance} relative to the Sun 
(where Fe/H is the ratio of the number of iron atoms to the number 
of hydrogen atoms in a volume unit of the stellar atmosphere). 
The sources of [Fe/H] data are: Giridhar (\cite{G83}), Fry \& Carney 
(\cite{FC97}), Groenewegen~et~al. (\cite{GRP04}), Andrievsky et~al. 
(\cite{Aetal02a}; \cite{Aetal02b}; \cite{Aetal02c}), Luck~et~al. 
(\cite{Letal03}), Andrievsky et~al. (\cite{Aetal04}; \cite{Aetal05}), 
Kovtyukh~et~al. (\cite{Ketal05a}, \cite{Ketal05b}), Romaniello~et~al. 
(\cite{Retal05}), Mottini (\cite{M06}), Yong~et~al. (\cite{Yetal06}), 
and Lemasle~et~al. (\cite{Lem07}). 

There are 187 Galactic Cepheids in our catalogue with known 
spectroscopic [Fe/H] values. Quite a few bright southern Cepheids, 
some of which are binary systems, (e.g., AX~Cir, V636~Sco) 
were neglected spectroscopically.

Various authors accept different solar chemical compositions. 
To homogenize the scale of [Fe/H] values, data 
were shifted to a common solar metallicity, 
$\log [n({\rm Fe})]=7.45$ on a scale where $\log [n({\rm H})]=12$ 
(Grevesse~et~al. \cite{Getal07}). Most [Fe/H] data have been taken from 
Andrievsky and his collaborators' papers (Andrievsky et~al. \cite{Aetal02a}; \cite{Aetal02b}; \cite{Aetal02c}; Luck~et~al. \cite{Letal03}, 
Andrievsky et~al. \cite{Aetal04}; \cite{Aetal05}; Kovtyukh~et~al. 
\cite{Ketal05a}; \cite{Ketal05b}). Their scale was shifted 
by 0.05 because they used an earlier [Fe/H] value for the solar chemical 
composition (Grevesse~et~al. \cite{Gretal96}). The [Fe/H] values obtained 
by others were transformed to this modified Andrievsky scale based on 
common Cepheids in the respective projects, Fry \& Carney (\cite{FC97}), 
Lemasle et~al. (\cite{Lem07}), Mottini (\cite{M06}), and Romaniello~et~al. 
(\cite{Retal05}). 
The transformation of [Fe/H] values obtained by Yong~et~al. (\cite{Yetal06}) 
to the common scale was taken from Luck~et~al. (\cite{Letal06}). 
The transformation equations are as follows: 
\begin{equation}
{\rm [Fe/H]}_{\rm And.} = 0.831 (\pm 0.233) \times {\rm [Fe/H]}_{\rm Fry} + 
0.053 (\pm 0.032),
\label{tr_fry}
\end{equation}
\begin{equation}
{\rm [Fe/H]}_{\rm And.} = 0.838 (\pm 0.196) \times {\rm [Fe/H]}_{\rm Lem.} + 
0.050 (\pm 0.030),
\label{tr_lem}
\end{equation}
\begin{equation}
{\rm [Fe/H]}_{\rm And.} =  0.627 (\pm 0.132) \times {\rm [Fe/H]}_{\rm Mot.} + 
0.013 (\pm 0.014),
\label{tr_mot}
\end{equation}
\begin{equation}
{\rm [Fe/H]}_{\rm And.} = 1.254 (\pm 0.291) \times {\rm [Fe/H]}_{\rm Rom.} + 
0.076 (\pm 0.039), {\rm and}
\label{tr_rom}
\end{equation}
\begin{equation}
{\rm [Fe/H]}_{\rm And.} = 0.965 (\pm 0.106) \times {\rm [Fe/H]}_{\rm Yong} + 
0.175 (\pm 0.130).
\label{tr_yon}
\end{equation}
These transformed [Fe/H] values were used only if no data were available 
from the databases of Andrievsky and his collaborators, to keep the 
data sample as homogeneous as possible. If Andrievsky et~al. published 
more than one [Fe/H] value for the same Cepheid, priority was given 
to the most recent value. The metallicity data of Giridhar (\cite{G83}) 
could not be transformed because of insufficient common stars. 

\subsubsection{Binarity status}
\label{sssect_binarity}

The presence of a companion may affect the photometric amplitudes
since an additional constant source of light always reduces the observable
amplitude of the brightness variation. The amount of amplitude decrease
is a function of the temperature and brightness differences between the
Cepheid and its companion(s), and depends also on the photometric band
considered. Physical and optical (i.e., line-of-sight) companions
are identical in this respect.

The situation is different for radial velocity variations. 
If a Cepheid belongs to a spectroscopic binary system, 
orbital and pulsational radial velocity changes are superimposed 
on each other. The observable radial velocity amplitude is, therefore, 
larger than the amplitude caused by the pulsational motion alone. 
If the spectroscopic orbit of the Cepheid is known, the amplitude of
pulsational radial velocity variations can be determined by removing
the orbital effect from the observed changes in the radial velocity.

Binarity is an important factor when studying the observable
amplitudes of Cepheids because more than 50\% of Galactic Cepheids 
belong to binary or multiple systems (Szabados \cite{Sz03b}).
A number of Cepheids may have undetected companions because of 
a selection effect preventing the discovery of
duplicity in fainter Cepheids (Szabados \cite{Sz03c}).

Binarity status assigned to individual Cepheids in Table~1 
is based on the online database of binary Cepheids\footnote{http://www.konkoly.hu/CEP/intro.html}, 
also giving references on star-by-star basis.

\subsubsection{Pulsation mode}
\label{sssect_mode}

Although the excited mode of the pulsation is a fundamental property 
of individual Cepheids, there are no infallible methods for
its determination. Cepheids in the Magellanic Clouds demonstrate that 
a separate $P$-$L$ relation exists for each pulsation mode 
(see e.g., Udalski et~al. \cite{Uetal99b}).
Cepheids pulsating in the first overtone are more luminous than 
fundamental mode oscillators of the same pulsation period, and
monoperiodic second overtone Cepheids are even more luminous.
Identification of the pulsation mode is necessary to determine
the luminosity of a given Galactic Cepheid but contradictory 
results are often found in the literature.

Phenomenologically, monoperiodic Cepheids can be divided into two groups. 
The majority of Cepheids have a large amplitude (larger than 0\fm5 in the 
Johnson $V$ photometric band) and an asymmetric light curve described by the 
well known Hertzsprung progression (Hertzsprung \cite{H26}). Members of the 
other group, containing Cepheids of low amplitude (smaller than 0\fm5 in 
Johnson $V$ band), are often referred to as s-Cepheids because their light 
curves are {\em s}inusoidal, {\em s}ymmetric, and of {\em s}mall amplitude.

For another type of radially pulsating stars, 
the RR~Lyrae variables, the pulsation mode can be inferred from
the shape and amplitude of the light curve. Variables of RRab subtype 
(asymmetric light curve of large amplitude) are fundamental mode pulsators, 
while the small amplitude RR~Lyraes with sinusoidal light curves 
(RRc subtype) pulsate in the first overtone (Castellani et~al. 
\cite{CCC03} and references therein).

Based on the study of Cepheids in the Large Magellanic Cloud (LMC), 
Connolly (\cite{C80}) suggested that small amplitude Cepheids with 
sinusoidal light curves are first overtone pulsators. 
Later on, this statement was generalized and, by an analogy with the 
RRab--RRc dichotomy, it was assumed that s-Cepheids in our Galaxy 
are also overtone pulsators. 
Editors of the General Catalogue of Variable Stars (Kholopov \cite{K85}) 
avoid firm statement in this respect: they mention that DCEPS stars 
(the GCVS type of s-Cepheids) are {\em possibly} first overtone pulsators 
and/or cross the instability strip for the first time after 
evolving off the main sequence.

Nowadays the mode determination is usually based on Fourier decomposition
of the light variations. Antonello et~al. (\cite{APR90}) found 
two suitable criteria that can be applied for discriminating s-Cepheids 
from their normal amplitude siblings. In the $R_{21}$ versus period diagram 
s-Cepheids form a lower sequence ($R_{21}<0.2$) below the normal amplitude 
Cepheids, while in the $\phi_{31}$ versus period diagram, they form an upper 
sequence ($\phi_{31}>3$) above the locus of normal amplitude Cepheids.

This quantitative procedure is usually followed by a step that is 
unjustified for Cepheids, the assumption that DCEPS stars and Cepheids 
pulsating in the first overtone mutually correspond to each other. 
Szab\'o et~al. (\cite{Szetal07}) discussed how low pulsational amplitudes 
do not necessarily relate to oscillations in an overtone. 
Their nonlinear pulsation models indicate that fundamental mode Cepheids 
of periods longer than 10 days have small amplitude oscillations 
near both edges of the instability strip. Cepheids in the Magellanic Clouds 
whose pulsation mode can be identified from the colour-magnitude diagram 
clearly demonstrate that there are s-Cepheids that oscillate in the 
fundamental mode and large amplitude Cepheids that exhibit first overtone 
pulsation (Udalski et~al. \cite{Uetal99a}, \cite{Uetal99b}).

In Table~1, the pulsation modes are  taken mostly from the extensive and 
homogeneous list compiled by Groenewegen \& Oudmaijer (\cite{GO00}).


\section{Period-amplitude relationships}
\label{sect_per-amp}

In his exhaustive paper on the $P$-$A$ relationship of Galactic Cepheids, 
Efremov (\cite{Ef68}) described how the period-dependent maximum amplitude 
is manifested in two local maxima at both $\log P = 0.73$ and 1.4, 
while the largest possible amplitude drops at $\log P = 0.96$; this drop 
is the consequence of a resonance between the fundamental eigenmode and 
its second overtone (Buchler et~al. \cite{Betal90}).
Long-period Cepheids tend to pulsate with larger amplitude than short period 
ones, but there is a range of possible amplitudes at each period. 
A lower boundary appears for normal amplitude Cepheids at $A_B=0.7$ magnitude. 
There is a separate group of small amplitude Cepheids, identical to the 
s-Cepheids, whose amplitude in the $B$ band is smaller than 0\fm6. Although 
Efremov (\cite{Ef68}) considered 0\fm4 to be the lower limit to the 
amplitudes of s-Cepheids in the $B$ band, discovery of even lower amplitude 
s-Cepheids (e.g., V636~Cas, BG~Cru, V1334~Cyg, V1726~Cyg, V440~Per) 
demonstrated that there is no lower amplitude limit for the pulsation of 
such Cepheids. 

\begin{figure*}[!]
\begin{center}
\includegraphics[width=12.5cm]{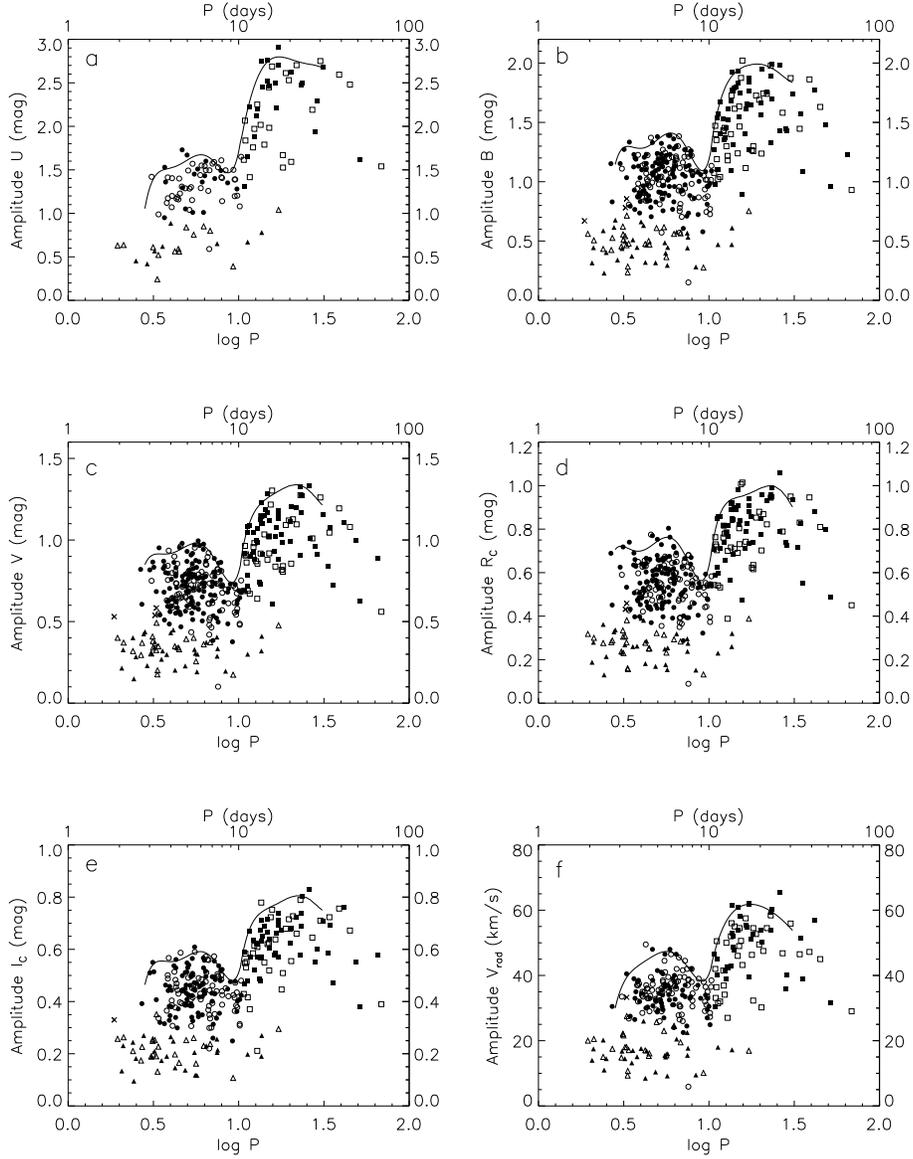}
\end{center}
\vspace*{-7mm}
\caption{Period-amplitude diagrams. Amplitudes in $U$, $B$, $V$, $R_{\rm C}$, 
and $I_{\rm C}$ photometric bands (panels a-e, respectively), 
and of radial velocity variations (panel f) are plotted. 
Circles and squares refer to fundamental mode Cepheids pulsating
with short ($\log P < 1.02$) and long period ($\log P > 1.02$), respectively, 
while triangles represent Cepheids pulsating in the first overtone. 
Filled symbols are used for Cepheids without known companion, 
empty symbols represent Cepheids belonging to binary (or multiple) systems. 
$\times$ symbols are used if the pulsation mode of the star is ambiguous. 
The upper envelope (for its construction see the text) 
is also shown in each plot.}
\label{fig_p-a}
\end{figure*}

The new $P$-$A$ graphs based on the data listed in Table~1 are shown 
in Figs.~\ref{fig_p-a}a-f for the $U$, $B$, $V$, $R_{\rm C}$, and 
$I_{\rm C}$ band and radial velocity amplitudes, respectively. 
In this Figure, circles denote short period ($\log P < 1.02$) Cepheids 
pulsating in the fundamental mode, squares refer to their long period 
counterparts ($\log P > 1.02$), triangles correspond to first overtone 
Cepheids, while $\times$ symbols are used to represent ambiguous 
pulsation mode. Empty symbols refer to Cepheids with known companion(s), 
filled symbols mean that there is no evidence of a companion. 

\begin{figure}[h]
\begin{center}
\includegraphics[width=8cm]{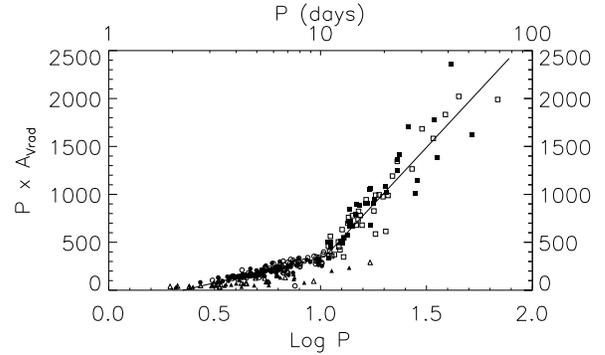}
\end{center}
\vspace*{-5mm}
\caption{Different behaviour of short- and long-period Cepheids. 
Meaning of the symbols is the same as for Fig.~\ref{fig_p-a}.
The best-fit linear sections intersect at $\log{P}=1.02$\,. 
s-Cepheids were not involved in the fitting procedure.}
\label{fig_dichotomy}
\end{figure}

The division between short- and long-period Cepheids is defined to be 
at $\log P = 1.02$, instead of $\log P = 0.96$  (Efremov \cite{Ef68}), 
or to be the pulsation period of 10 days adopted in studies dealing with 
the break in the $P$-$L$ relationship (e.g. Sandage et~al. \cite{Sandetal09}). 
This choice of a longer period limit (which corresponds to $P = 10\fd47$) 
is supported by Fig.~\ref{fig_dichotomy}. In this diagram, the product 
of the pulsation period and the radial velocity amplitude is plotted 
against $\log P$. This product has no direct physical meaning but, 
due to its dimensions, it is related to the variation in the stellar 
radius during a pulsational cycle. The different behaviour of short- and 
long-period Cepheids (and the third group, namely the s-Cepheids) is obvious.
The intersection of the two linear sections fitted to
the data representing fundamental mode Cepheids indicates that 
the break occurs at $\log P = 1.02$. 

To determine the upper envelope to our $P$-$A$ plots, we followed the 
statistically sound method applied by Eichendorf \& Reinhardt (\cite{ER77}) 
with some modifications. We divided the range of $\log{P}$ into equal 
intervals. Each interval has a width of 0.05 in $\log{P}$. The problem 
of individual bins containing different numbers of Cepheids can be mitigated 
by weighting the envelope-points by the number of stars in the given bin.

The largest observed amplitude in each interval was taken as a
preliminary envelope point. The corresponding period for 
each envelope point was calculated to be the mean $\log P$ 
of Cepheids in the given interval. Unlike Eichendorf \& Reinhardt's 
(\cite{ER77}) procedure, we divided the envelope into two parts 
and the fits were determined separately. The two parts cover 
$0.4 < \log{P} < 1.02$ and $1.02 < \log{P} < 1.5$ intervals, representing 
short- and long-period Cepheids, respectively. Both the decrease in the 
pulsation amplitude near $\log{P}=1.0$, and the different behaviour of 
short- and long-period Cepheids justify this division. Outside the 
$0.4 < \log{P} < 1.5$ interval, the sample contains an insufficient 
number of Cepheids.

The most realistic upper envelopes were obtained by 
a least squares fit to the preliminary envelope points 
for each part in the form of a fifth order polynomial
\begin{equation}
{\rm Envelope} = c_0 + \sum_{i=1}^5 c_i\times(\log{P})^i
\label{polyn}
\end{equation}
Coefficients describing the upper envelopes and their errors for the 
amplitudes in $U$, $B$, $V$, $R_{\rm C}$, $I_{\rm C}$ bands
and for radial velocity amplitudes are listed in Table~\ref{env_coeff}.
These upper envelopes represent the largest possible pulsation 
amplitude at a given period. A few datapoints above the upper envelope 
in Fig.~\ref{fig_p-a} indicate the uncertainties in the envelope curves. 
Nevertheless, the fitting could not have been constrained such that the 
envelope passed through the points of the largest true amplitudes.

\begin{table*}[!]
\caption{Coefficients of the envelope curves (with the formal errors given in
parentheses below the respective coefficient)}
\begin{tabular}{lr@{\hskip1mm}r@{\hskip1mm}r@{\hskip1mm}r@{\hskip1mm}r@{\hskip1mm}rcr@{\hskip1mm}r@{\hskip1mm}r@{\hskip1mm}r@{\hskip1mm}r@{\hskip1mm}r}
\hline
\noalign{\smallskip}
 & $c_0$ & $c_1$ & $c_2$ & $c_3$ &  $c_4$ & $c_5$ & & $c_0$ & $c_1$ & $c_2$ & $c_3$ & $c_4$ & $c_5$\\
\noalign{\smallskip}
\cline{2-7} \cline{9-14}
\noalign{\smallskip}
& \multicolumn{6}{c}{$\log P < 1.02$} & & \multicolumn{6}{c}{$\log P > 1.02$} \\
\noalign{\smallskip}
\hline
\noalign{\smallskip}
$U$ & $-$126.5 & 936.3 & $-$2712.3 & 3882.8 & $-$2742.4 & 763.8 & & 241.7 & $-$1209.8 & 2267.4 & $-$2024.2 & 873.9 & $-$147.3\\
& (75.6) & (525.7) & (1439.9) & (1941.5) & (1290.6) & (338.7) & & (1538.1) & (6254.0) & (10110.7) & (8132.8) & (3251.2) & (517.7) \\[0.8ex]
$B$ & $-$95.4 & 726.8 & $-$2151.1 & 3128.6 & $-$2233.6 & 625.9 & & $-$979.8 & 3783.1 & $-$5836.3 & 4503.0  & $-$1736.4 & 267.5 \\
& (26.2) & (189.8) & (539.1) & (752.3) & (516.3) & (139.5) & & (2003.3) & (8383.5) & (13983.4) & (11612.6) &(4796.2) & (789.7) \\[0.8ex]
$V$ & $-$46.7 & 364.5 & $-$1099.0 & 1629.0 & $-$1184.9 & 337.9 & & $-$885.1 & 3461.5 & $-$5396.1 & 4194.3 & $-$1624.5 & 250.7 \\
& (12.3) & (91.2) & (265.1) & (378.6) & (265.7) & (73.3) & & (1594.3) & (6656.8)& (11072.6) & (9166.6) & (3778.5) & (620.2) \\[0.8ex]
$R_{\rm C}$ & $-$37.8 & 302.2 & $-$928.7 & 1396.9 & $-$1027.7 & 295.7 & & $-$750.6 & 2902.9 & $-$4466.9 & 3420.9 & $-$1303.2 &  197.5\\
& (12.8) & (94.2) & (272.7) & (386.7) & (269.2) & (73.6) & & (505.6) & (2050.6) & (3308.8) & (2656.5) &(1060.7)& (168.5) \\[0.8ex]
$I_{\rm C}$ & $-$37.6 & 282.2 & $-$822.3 & 1179.8 & $-$832.7 & 231.1 & & $-$579.8 & 2266.9 & $-$3531.1 & 2741.5 & $-$1060.3 & 163.4 \\
& (9.0) & (65.9) & (188.2) & (264.3) & (182.4) & (49.5) & & (505.5) & (2075.3) & (3390.8) & (2756.8) & (1114.6)& (179.2) \\[0.8ex]
$V_{\rm RAD}$ & $-$2074 & 15367 & $-$44551 & 64191 & $-$45746 & 12855 & & $-$11217 & 41017 &$-$59974 & 44053 & $-$16236 & 2397 \\
& (2083) & (14664) & (40690) & (55672) & (37584) & (10010) & & (41150) & (167648) & (271723) & (219110) & (87922) & (14035) \\
\noalign{\smallskip}
\hline
\end{tabular}
\label{env_coeff}
\end{table*}


All individual graphs in Fig.~\ref{fig_p-a}a-f exhibit similar patterns, 
as far as the shape of the upper envelope and the dichotomy between 
normal and small amplitude Cepheids are concerned, independently of 
the wavelength of the photometric band and even for the amplitude 
of the $V_{\rm RAD}$ variations. The upper envelopes define the largest 
pulsational amplitudes at $\log P = 0.76$ (in accordance with the value 
given by Efremov \cite{Ef68}) and at $\log P = 1.30$ (differing from 
Efremov's corresponding value of 1.4). The minimum amplitude 
at intermediate periods predicted by the envelope curves occurs
at $\log P = 0.96$ in perfect agreement with the value given by Efremov. 
Nevertheless, we divide the normal amplitude Cepheids into 
two groups at the period limit of $\log P = 1.02$, in accordance with
the dichotomy pointed out earlier in this Section.

Fundamental mode Cepheids belonging to binary systems tend to have 
smaller photometric amplitudes than their solitary counterparts,
but this effect is not discernible in the case of s-Cepheids. 
Because the majority of companions to Cepheids are blue stars, 
the brightness difference between the Cepheid and its companion 
usually decreases towards shorter wavelengths, thus the observable 
amplitudes in $U$ and $B$ bands are lower than for the 
$V$ and  $R$ bands. Numerical data listed in Table~\ref{aveamp},
i.e., the average amplitudes for each mode of oscillation and each 
photometric band studied, and for radial velocity variations, 
separately for binary and solitary Cepheids, support these statements. 
The average $A_U$ of Cepheids with companions is about 85\% of 
the corresponding value for solitary Cepheids, and this ratio decreases 
to between 0.92 and 0.94 for amplitudes in other photometric bands. 
In contrast, observed pulsational radial velocity amplitudes should not 
depend on the duplicity status of Cepheids, and this is confirmed 
by the observed values: the average pulsational radial velocity amplitude 
of Cepheids with companions differs by only 4\% from the corresponding 
amplitude of solitary Cepheids. 

\begin{table*}[thb]
\caption{Average pulsational amplitudes}
\begin{tabular}{l@{\hskip0mm}c@{\hskip1mm}c@{\hskip1.5mm}c@{\hskip2.5mm}c@{\hskip1mm}c@{\hskip1.5mm}c@{\hskip2.5mm}
c@{\hskip1mm}c@{\hskip1.5mm}c@{\hskip2.5mm}c@{\hskip1mm}c@{\hskip1.5mm}c@{\hskip2.5mm}c@{\hskip1mm}c@{\hskip1.5mm}
c@{\hskip2.5mm}c@{\hskip1mm}c@{\hskip1mm}c}
\hline
\noalign{\smallskip}
{\em Mode} \&  & $A_U$ &  $\sigma_{A_U}$ & $n_U$ & $A_B$ & $\sigma_{A_B}$ & $n_B$ & $A_V$ & $\sigma_{A_V}$ & $n_V$ & $A_{R_C}$ & $\sigma_{A_R}$ & $n_R$ & $A_{I_C}$ & $\sigma_{A_I}$ & $n_I$ & $A_{V_{\rm RAD}}$ & $\sigma_{A_{V_{\rm RAD}}}$ & $n_{V_{\rm RAD}}$\\
\noalign{\smallskip}
\cline{2-3} \cline{5-6} \cline{8-9} \cline{11-12} \cline{14-15} \cline{17-18} 
\noalign{\smallskip}
Binarity   & \multicolumn{2}{c}{(m)} &  & \multicolumn{2}{c}{(m)} &  & \multicolumn{2}{c}{(m)} &  & \multicolumn{2}{c}{(m)} &  & \multicolumn{2}{c}{(m)} &  & \multicolumn{2}{c}{(km/s)} &   \\
\noalign{\smallskip}
\hline
\noalign{\smallskip}
{\em Fundamental} &&&&&&&&&&&&&&&&&&  \\
Solitary   &  1.882 & 0.587 & 40 & 1.193 & 0.320 & 175 & 0.810 & 0.205 & 189 & 0.635 & 0.161 & 176 & 0.507 & 0.120 & 141 & 39.56 & 10.19 & 101 \\
Binary     &  1.585 & 0.476 & 67 & 1.155 & 0.308 & 117 & 0.772 & 0.196 & 120 & 0.606 & 0.155 & 116 & 0.483 & 0.122 &  109 & 38.01 &  8.38 & 118 \\
{\em First overtone}  &&&&&&&&&&&&&&&&&& \\
Solitary   &  0.631 & 0.155 &  8 & 0.462 & 0.120 &  30 & 0.314 & 0.081 &  30 & 0.257 & 0.070 &  28 & 0.197 & 0.051 &  24 & 15.76 &   4.07 &  26 \\
Binary     &  0.644 & 0.202 & 14 & 0.500 & 0.124 &  25 & 0.342 & 0.078 &  25 & 0.275 & 0.063 &  24 & 0.221 & 0.054 &  23 & 18.24 &  4.68  & 24 \\
\noalign{\smallskip}
\hline
\noalign{\smallskip}
\end{tabular}
\label{aveamp}
\end{table*}

The absence of the photometric effect of companions on the amplitude of 
first overtone Cepheids is somewhat surprising. Although their sample 
is smaller, the general behaviour cannot be doubted: 
binaries pulsate with a larger (by a factor of about 1.1) amplitude 
than overtone Cepheids without known companions (see Table~\ref{aveamp}).
This behaviour implies that first overtone Cepheids can oscillate with 
any amplitude smaller than the largest possible value, while Cepheids 
pulsating in the fundamental mode prefer oscillating with a large amplitude, 
even though their physical properties do not place them in the middle of 
the instability strip. A remarkable exception is \object{V440~Persei}, 
which is classified as a fundamental mode Cepheid in spite of its 
extremely low amplitude (Szab\'o et~al.~\cite{Szetal07}).

Because companion stars leave the amplitude of radial velocity 
variations unchanged (having removed the orbital effect), the $P$-$A$ diagram 
constructed for radial velocity data is expected to show a relatively smaller 
spread than the period -- photometric amplitude graph 
in Fig.~\ref{fig_p-a}. In contrast to this expectation, the pattern of
the $A_{V_{\rm RAD}}$ versus $\log P$ graph for Galactic Cepheids is largely 
similar to that of photometric amplitude versus $\log P$ diagrams without 
a noticeable decrease in the ratio of minimum and maximum amplitudes 
at a given pulsation period (and treating s-Cepheids separately). 
This feature can be explained, at least in part, by the lower precision 
of the radial velocity data and possibly by unrecognised spectroscopic companions. 
In this latter case, the observable amplitude is larger than the pulsational 
amplitude. The excess is caused by the contribution from the projected orbital 
motion superimposed on the amplitude of pulsational origin. An additional 
cause of the wide range of observed radial velocity amplitudes is the effect 
of the atmospheric metal content to be discussed in Paper~II.

\section{Discussion}
\label{sect_disc}

The observable amplitude of a Cepheid may depend on: the effective temperature 
of the star, as well as its luminosity (Sandage et~al. \cite{Sandetal04}), 
which are both related to the position of the star within the instability 
strip; the atmospheric metallicity (by means of effect on the energy balance of 
the pulsation); helium content; and the presence of companions.
Investigations of the effects of the temperature, luminosity (which correlates 
with surface gravity in the case of Cepheids), and chemical composition 
on the oscillation amplitude were beyond the scope of this paper. 
Here we have concentrated on relations between various amplitudes, 
including their period dependence. Interrelations of various amplitudes
facilitate the identification of binary stars among Cepheids. In view of the photometric effects of companions, known binaries have to be excluded when studying the intrinsic pulsational behaviour of the amplitudes. 

Binarity is usually identified by means of spectroscopy. Owing to the 
regularity of the pulsation, there are also photometric methods for detecting 
the duplicity of Cepheids. Because of the finite range of amplitudes of 
solitary Cepheids, small or moderate amplitudes do not necessarily 
hint at the presence of a companion. Properly selected combinations of 
photometric amplitudes in different colours, however, can be suitable 
duplicity indicators. Photometric duplicity tests are based on the 
amplitudes in various bands and colour indices, and their phase relations 
(see the summary in Szabados \cite{Sz03a}).

\subsection{Ratio of amplitudes}
\label{ssect_inter}

\begin{figure*}[!]
\begin{center}
\includegraphics[width=12.5cm]{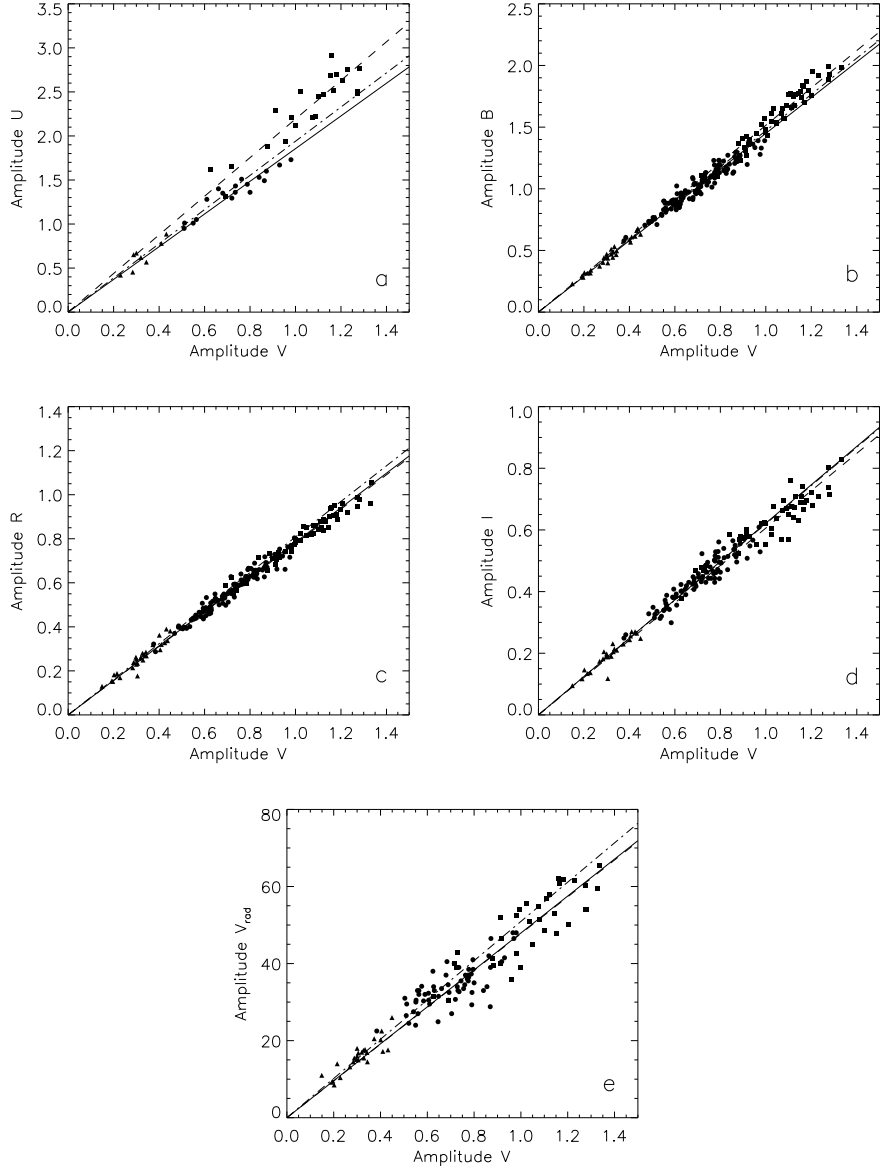}
\end{center}
\vspace*{-5mm}
\caption{Dependence of $A_U$, $A_B$, $A_R$, $A_I$, and $A_{V_{\rm RAD}}$ 
on $A_V$. Circles denote short period ($\log P < 1.02$) 
Cepheids pulsating in the fundamental mode, squares refer to their 
long-period counterparts ($\log P > 1.02$), triangles represent overtone pulsators. Linear least squares fits are also shown: a dashed line 
for long-period Cepheids, a solid line for short-period Cepheids, 
and a dash-dotted line for Cepheids pulsating in the first overtone. 
These plots only involve Cepheids without known companions.}
\label{amp_amp}
\end{figure*}

Amplitudes observed in two photometric bands are normally tightly 
correlated. During the first detailed study of this correlation, 
van~Genderen (\cite{vG74}) noted that the average ratio of 
$A_V/A_B$ differs for the largest amplitude Cepheids 
from those pulsating with an amplitude smaller than 1\fm5 in the $B$ band. 
He obtained $A_V = 0.64 A_B$ for $A_B > 1\fm5$, while the complete 
$A_V$ versus $A_B$ plot could be approximated by a line of slope 0.67.
Because only long-period Cepheids can pulsate with $A_B > 1\fm5$, 
this dichotomy means a difference between the pulsations of 
short- and long-period Cepheids.

In the case of classical Cepheids, photometric amplitudes decrease 
with increasing wavelength: 
$A_{{\lambda}_1}/A_{{\lambda}_2} > 1$ (${\lambda}_1 < {\lambda}_2$). 
Freedman (\cite{F88}) determined the average ratios to be 
$A_B : A_V : A_{R_{\rm J}} : A_{I_{\rm J}} = 1.00:0.67:0.44:0.34$ from
photoelectric observations of 20 classical Cepheids published by Wisniewski 
\& Johnson (\cite{WJ68}). However, 16 Cepheids in that sample belong to 
binary systems, which obviously falsifies the derived amplitude ratios.

When considering $A_U$ and $A_B$ amplitudes, there are exceptions: 
the $A_U/A_B$ ratio is about unity for V495~Cyg, V1334~Cyg 
(a known spectroscopic binary), and \object{V950~Sco}, indicating a 
bright blue companion to these Cepheids. For V950~Sco, this is 
the first evidence of binarity, while for V495~Cyg this ratio is a 
further piece of evidence in addition to the large $q$ value mentioned 
in Sect.~\ref{ssect_ar}. 

The $A_V/A_B$ ratio of V495~Cyg (0.737) also indicates a blue companion, 
so does the large $A_V/A_B$ value (0.722) of UZ~Cas. 
There are a number of Cepheids whose hot companion 
has been revealed by ultraviolet spectroscopy (see the online database 
on Cepheids in binaries and its description by Szabados \cite{Sz03b}). 
The excessive $A_V/A_B$ ratios of these Cepheids, e.g., RW~Cam (0.751), 
KN~Cen (0.723), SU~Cyg (0.767), S~Mus (0.728), and AW~Per (0.714) 
also confirm the diagnostic value of this amplitude ratio 
in detecting blue companions.

Figure~\ref{amp_amp} is a collection of amplitude-amplitude dependences. 
Panels (a) through to (e) show the $A_U$, $A_B$, $A_R$, $A_I$, 
and $A_{V_{\rm RAD}}$ values as a function of $A_V$, respectively. 
Because companion stars exert a wavelength dependent influence on 
the photometric amplitudes, only Cepheids classified as `solitary' 
have been taken into account in constructing Fig.~\ref{amp_amp}. 

Remarkably, this cleaned sample shows a dichotomic behaviour: 
long-period Cepheids ($\log P > 1.02$; filled squares) are 
characterized by a different slope in the $A_{\lambda}$ versus $A_V$ 
relationship than Cepheids pulsating with short periods 
($\log P < 1.02$; filled circles). This behaviour is in accordance 
with the finding of van~Genderen (\cite{vG74}) but now the dichotomy 
in amplitude ratios is generalized to other photometric bands.

The slopes of the linear fits (forced to cross the origin in each panel of 
Fig.~\ref{amp_amp}) are summarized in Table~\ref{amp-amp_sum}, 
which also includes the slopes of other relationships not shown 
in Fig.~\ref{amp_amp}. 
Column~6 of Table~\ref{amp-amp_sum} lists the ratio of the slopes obtained 
for long period over the slope for short period pulsators. Marked as 
triangles in Fig.~\ref{amp_amp}, s-Cepheids, behave in a similar way 
to short-period Cepheids.

\begin{table*}[thb]
\caption{Slopes of the linear relationships between various amplitudes 
(omitting the known binaries). `Ratio' refers to the rate of slopes
in the sense value for long- over value for short-period Cepheids.}
\begin{tabular}{l@{\hskip5mm}r@{\hskip2mm}l@{\hskip2mm}r@{\hskip5mm}r@{\hskip2mm}l@{\hskip2mm}r@{\hskip5mm}r@{\hskip7mm}r@{\hskip2mm}l@{\hskip2mm}r}
\hline
\noalign{\smallskip}
 & Slope & $\sigma_{\rm slope}$ & N & Slope & $\sigma_{\rm slope}$ & N & Ratio & Slope & $\sigma_{\rm slope}$ & N\\
\noalign{\smallskip}
\cline{2-7} \cline{9-11}
\noalign{\smallskip}
Amplitudes & \multicolumn{7}{c}{Fundamental mode} & \multicolumn{2}{c}{1st overtone}\\
\noalign{\smallskip}
involved & \multicolumn{3}{c}{$\log P < 1.02$} & \multicolumn{3}{c}{$\log P > 1.02$} & & \\
\hline
\noalign{\smallskip}
$A_U$\,-\,$A_V$ & 1.856 & 0.027 & 18 & 2.192 & 0.037 & 22 & {\em 1.182} & 1.940 & 0.045 & 8 \\
$A_B$\,-\,$A_V$ & 1.449 & 0.007 & 119 & 1.515 & 0.008 & 54 & {\em 1.046} & 1.471 & 0.007 & 30 \\
$A_R$\,-\,$A_V$ & 0.784 & 0.003 & 118 & 0.779 & 0.004 & 56 & {\em 0.994} & 0.808 & 0.005 & 28 \\
$A_I$\,-\,$A_V$ & 0.622 & 0.004 & 90 & 0.606 & 0.005 & 49 & {\em 0.974} & 0.620 & 0.003 & 24 \\
$A_{V_{\rm RAD}}$\,-\,$A_V$ & 47.93 & 0.73 & 64 & 47.76 & 0.86 & 35 & {\em 0.996} & 50.93 & 0.90 & 26 \\
$A_R$\,-\,$A_B$ & 0.540 & 0.002 & 111 & 0.513 & 0.004 & 51 & 0.950 & 0.548 & 0.008 & 28 \\
$A_I$\,-\,$A_B$ & 0.424 & 0.003 & 82 & 0.398 & 0.004 & 46 & 0.939 & 0.423 & 0.007 & 24 \\
$A_{V_{\rm RAD}}$\,-\,$A_B$ & 32.81 & 0.50 & 63 & 31.12 & 0.59 & 33 & 0.948 & 34.62 & 0.92 & 26\\
\noalign{\smallskip}
\hline
\end{tabular}
\label{amp-amp_sum}
\end{table*}

The ratio of the slopes itself is also wavelength dependent: the larger 
the wavelength difference of the photometric bands involved, the larger 
is the difference in the slopes for the long- and short-period Cepheids.

We checked by statistical tests whether fits to short- and long-period 
Cepheids really have different slopes in the $A_{\lambda}$ versus $A_V$ 
diagram, or a fit with a single straight line or a parabola is more 
appropriate. The Student's two-sample t-test shows whether the means 
of the two subsamples (short and long period stars) are different. 
This test was used assuming that the two distributions have the same variance 
(to be checked with an F-test). If the points correspond to the fit, 
the means of the subsamples should be the same (null hypothesis).

In our case, the variances were the same with a confidence range 
of 95\%, so we could use the Student's two-sample t-test. 
Neither in the case of the single straight line, nor of the parabola 
was the null hypothesis acceptable with a confidence range of 
95\%, so the means of the subsamples are not the same, and 
a single linear fit is not consistent with the points.

An extrinsic cause of the different $A_B/A_V$ ratios of Cepheids
with short and long pulsation periods might be that the effective 
wavelength, $\lambda_{\rm eff}$, of the photometric bands depends on 
the effective temperature, $T_{\rm eff}$, of the observed star: 
$\lambda_{\rm eff}$ shifts to longer values for cooler stars,
which, in turn, causes a stronger decrease in $A_B$ than $A_V$ 
towards lower $T_{\rm eff}$, i.e., longer period Cepheids. This effect
would result in a continuously decreasing $A_B/A_V$ ratio toward 
longer pulsation periods. The observed effect is, however, just the 
opposite: the slope of the fitted line is larger for the sample with 
$\log{P}> 1.02$ than for the group of short-period Cepheids.
The behaviour of the longest period Cepheids also implies that the
dichotomy in the amplitude ratio has nothing to do with the
convolution of filter transmission with the spectral energy
distribution. Both \object{II~Car} and \object{GY~Sge} 
(disregarding long-period Cepheids with companions) are located
close to the locus of much shorter period (11-14 days) Cepheids 
of intermediate amplitudes in the $A_B$ versus $A_V$ diagram, 
instead of deviating upwards.

It is noteworthy that Coulson \& Caldwell (\cite{CC89}) found a linear
relationship between  $A_{\lambda}/A_V$  and $\log P$ that is valid for 
the $U$, $B$, and $I$ photometric bands. Amplitude ratios derived from 
our database, plotted in Fig.~\ref{fig_ar}, however, indicate that the 
linear fit is a rough approximation that also contradicts the dichotomic
behaviour evident in Fig.~\ref{amp_amp}.

A common feature seen in each panel of Fig.~\ref{fig_ar} is the
considerable scatter in the data points. A part of the scatter can be 
explained by the observed metallicity dependence of the amplitude ratios 
that will be discussed in Paper~II. Binary companions may also have 
an adverse effect on various ratios of photometric amplitudes depending on 
the temperature difference between the Cepheid and its companion. It is noteworthy that a majority of points strongly deviating upwards or downwards correspond to binaries (open symbols in Fig.~\ref{fig_ar}).
Some deviating points representing solitary Cepheids may indicate binarity. 
Spectroscopy and multicolour photometry of these variables (\object{FM~Car}, 
\object{BP~Cas}, \object{V459~Cyg}, \object{V924~Cyg}, \object{UY~Per}, 
\object{V773~Sgr} -- possible blue companion; \object{CY~Car}, \object{AY~Cen}, 
\object{GI~Cyg} -- possible red companion) are recommended.

Large deviations from the typical values of the $A_U/A_V$ and $A_B/A_V$ 
amplitude ratios of SU~Crucis are unique among Cepheids and cannot be 
explained by any reasonable companion star, although Coulson \& Caldwell 
(\cite{CC89}) and Laney \& Stobie (\cite{LS93}) hypothesized an extremely 
red companion.

\subsection{The $A_{V_{\rm RAD}}/A_B$ amplitude ratio}
\label{ssect_ar}

The ratio of the amplitudes of the radial velocity to photometric variations
is an indicator of the pulsation mode, based on purely observational data. 
The theoretical background behind why this ratio is suitable for 
differentiating between the modes of pulsation is laid by Balona \& Stobie (\cite{BS79}). Based on their linear pulsational model, one expects a 
1/0.7 = 1.43 times larger value of amplitude ratio for the first overtone pulsation compared with the fundamental mode oscillation (in a given photometric band) because the period ratio of the two excited modes 
is about 0.7 (see Eq.~\ref{fundamentalization}).

The opportunity for determining the pulsation mode from the
ratio of radial velocity to photometric amplitudes stimulated an
in-depth study of the $A_{V_{\rm RAD}}/A_B$ amplitude ratio.
The $A_B$ amplitude was chosen because its value is known 
for most Cepheids and it is larger than in other bands (except $U$, 
but $A_U$ values are available for a much smaller sample), 
therefore relative errors are smaller.
In what follows, the value of the $A_{V_{\rm RAD}}/A_B$ 
amplitude ratio is referred to as $q$.

Cepheids in the Magellanic Clouds are excellent test objects for checking 
the validity of theoretically predicted $q$ because their pulsation mode 
can be identified even in the case of single-mode oscillation. 
From observational data of 29 fundamental mode (F) and 9 first overtone 
(1OT) Magellanic Cepheids taken from the McMaster Cepheid Photometry and 
Radial Velocity Archive (Welch~\cite{W98}), the ratio of 
$q_{\rm F}/q_{\rm 1OT} = 0.72 \pm 0.17$ can be obtained. 
Although the ratio itself corresponds to the expectation, its precision 
is not satisfactory.

Cepheids oscillating simultaneously in two radial modes provide another 
test. The ratio of the $q$ values determined for the most well observed 
beat Cepheids, \object{TU~Cas} and \object{EW~Sct}, using the photometric data of Berdnikov (\cite{B08}) and radial velocity data of Gorynya~et~al. 
(\cite{Getal92, Getal96}) is $q_{\rm F}/q_{\rm 1OT} = 0.747 \pm 0.025$ 
(based on amplitudes listed in Table~\ref{beatamp}).  
When performing Fourier decomposition, linear combinations of the frequencies 
of the two excited modes ($f_0+f_1$, $2f_0+f_1$, $3f_0+f_1$, $f_1-f_0$, 
i.e., the most relevant coupling-terms present in the pulsation of the 
beat Cepheids) have also been taken into account.

Because companions reduce the observable photometric amplitudes, 
and an unrecognized orbital motion superimposed on pulsational 
changes results in an increased $A_{V_{\rm RAD}}$, a larger-than-normal 
value of $A_{V_{\rm RAD}}/A_B$ may indicate a companion. This effect of
companions lessens the diagnostic role of $q$ in assigning 
the mode of pulsation.

\begin{table}[]
\caption{Average values of the $q$ amplitude ratio}
\begin{tabular}{lccc}
\hline
\noalign{\smallskip}
Sample& $q$ & $\sigma$ & N \\
\noalign{\smallskip}
\hline
\noalign{\smallskip}
Fundamental-mode Cepheids (solitary) & 32.79 & 4.01 & 96 \\
\ \ \ \  $\log P < 1.0$ & 33.54 & 3.95 & 62 \\
\ \ \ \  $\log P > 1.0$  & 31.42 & 3.79 & 34 \\
Fundamental-mode Cepheids (binary) & 33.66 & 5.68 & 115 \\
\noalign{\smallskip}
\hline
\noalign{\smallskip}
First overtone Cepheids (solitary) & 35.23 & 4.99 & 26 \\
First overtone Cepheids (binary)   & 36.58 & 4.59 & 24 \\
\noalign{\smallskip}
\hline
\end{tabular}
\label{AR_sum}
\end{table}

\begin{figure*}[!]
\begin{center}
\includegraphics[width=11.5cm]{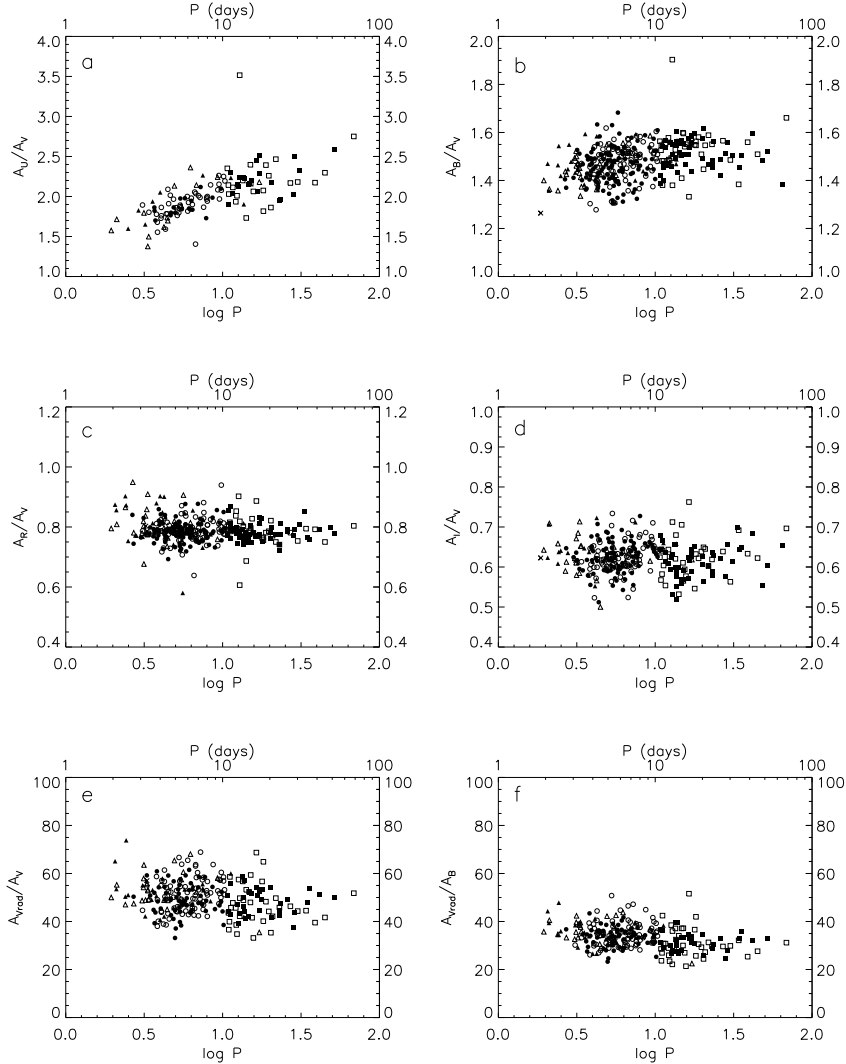}
\end{center}
\vspace*{-5mm}
\caption{Ratio of various amplitudes as a function of the pulsation period. 
The meaning of the symbols is the same as for Fig.~\ref{fig_p-a}. 
The effect of binarity on these amplitude ratios is discussed in the text. 
The strange outlier in the top of the two uppermost panels corresponds 
to SU~Cru.}
\label{fig_ar}
\end{figure*}

A linear relationship would be expected between the radial 
velocity and photometric amplitudes from the similar pattern of 
the respective $P$-$A$ plots (as seen in Fig.~\ref{fig_p-a}). 
In spite of the relatively low measurement errors, values 
plotted in Fig.~\ref{fig_ar}f have significant scatter: extrema may be 
20-30 per cent larger or smaller than the mean value of $q$ at any 
given period. These deviations from the mean greatly exceed the uncertainty 
in the determination of amplitudes. 
The reliability of the $q$ values determined from the observations 
can be estimated with the help of more than 20 Cepheids 
for which two values of $q$ could be calculated from independent data series. 
The average difference between the two independently obtained values 
is about 4\%. Therefore, the relative error in the $q$ derived from 
well covered photometric and radial velocity phase curves does not exceed 
$\pm 1.7$.

Cepheids pulsating in the first overtone occupy a region overlapping
that of fundamental pulsators at short periods, although one 
would expect their occurrence in distinct regions shifted vertically 
in panels $e$ and $f$ of Fig.~\ref{fig_ar}. Omitting known binaries,
an average ratio, $(q)_{\rm F}/(q)_{\rm 1OT} = 0.93 \pm 0.21$
is obtained, which deviates from the corresponding value obtained using 
double-mode and Magellanic Cepheids. 

A possible cause of the extraordinary ratio of the $q$ values is 
that some Cepheids classified as first overtone pulsators 
oscillate, in fact, in the fundamental mode.

Although binaries tend to have larger $q$ values on average than solitary 
Cepheids, a larger-than-average value of $q$ does not necessarily 
imply duplicity of the given variable, keeping in mind the width of 
the interval of normal $q$ values. This conclusion is confirmed numerically
by the data listed in Table~\ref{AR_sum} indicating marginally larger 
$q$ values for Cepheids in binaries than for their solitary counterparts.

For a given pulsation mode, an extremely large value of this ratio is a hint 
that the Cepheid may have a companion. According to this duplicity 
indicator, fundamental pulsators \object{UZ~Cas}, \object{VW~Cas}, 
\object{BP~Cas}, \object{CT~Cas}, and \object{V495~Cyg}, and the first 
overtone Cepheid \object{CR~Cep} certainly belong to binary systems. 
Further evidence of the duplicity of UZ~Cas and V495~Cyg is 
their relatively low value of the $A_B/A_V$ amplitude ratio indicating 
a blue companion (see Sect.~\ref{ssect_inter}). The binarity of VW~Cas, 
CR~Cep, and V495~Cyg is also suspected from their $m$ and/or $k$ parameters 
(see Sect.~\ref{ssect_slope} and Sect.~\ref{ssect_cawd}).
In view of these independent hints, UZ~Cas, VW~Cas, V495~Cyg, and CR~Cep are
considered as members of binaries throughout this paper.

The $q = 47.8$ value of \object{LR~TrA} indicates that this Cepheid
either has a companion or pulsates in the second overtone. 
Single-mode second-overtone Cepheids are known in the 
Small Magellanic Cloud (Udalski et~al.~\cite{Uetal99b})
but there is no straightforward method to identify them in our Galaxy.

\subsection{Wavelength dependence of amplitudes}
\label{ssect_wavedep}

It is reasonable to define a single parameter containing information 
about the wavelength dependence of the individual amplitudes 
and their ratios. This parameter is more sensitive to the presence of 
either blue or red companions than ratios of selected pairs of amplitudes.

\subsubsection{The $m$ parameter}
\label{ssect_slope}

Fernie (\cite{F79}) conceived the idea of plotting normalized photometric 
amplitudes versus the reciprocal of the effective wavelength of the given 
photometric band, which led him to the discovery of a hot companion to SU~Cyg. 

Amplitudes normalized to $A_B$ as a function of the reciprocal wavelength 
follow a linear pattern for a given Cepheid if photometric amplitudes observed 
in $U$, $B$, $V$, and $R$ bands are involved and the wavelength is expressed 
in $\mu$m (Szabados \cite{Sz00}) as 
\begin{equation}
A_{\lambda}/A_B = m\times(1/{\lambda}) + {\rm const.},
\label{eq-slope}
\end{equation}

\noindent where the slope of this line, $m$, infers that a companion
is present, in that a much larger slope than the average value indicates 
a red companion, while the presence of a blue companion corresponds to a shallower slope. We note that a different slope parameter was used in an 
earlier paper (Szabados \cite{Sz93}), which was defined to be the slope 
of the line fitted to the values of the $U$, $B$, $V$, and $R$ amplitudes 
plotted against the decimal logarithm of the wavelength of the 
relevant passband.

\begin{figure}
\begin{center}
\includegraphics[width=7cm]{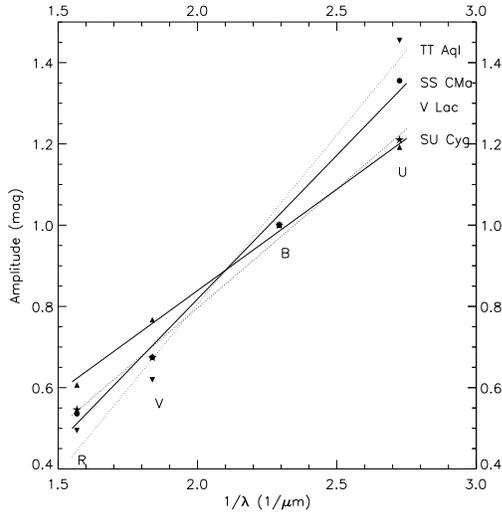}
\end{center}
\vspace*{-3mm}
\caption{Dependence of normalized photometric amplitudes on the 
reciprocal wavelength. The linear fit to the points representing individual 
Cepheids provides the definition of the $m$ parameter. 
Two pairs of Cepheids show the usefulness of $m$: 
SS~CMa (normalized amplitudes are marked by bullets) and SU~Cyg (triangles)
have blue companions (solid lines), while TT~Aql (triangle down) and V~Lac 
(star symbol) are solitary Cepheids (dotted lines) with pulsation period 
similar to that of SS~CMa and SU~Cyg, respectively.}
\label{fig_slope_def}
\end{figure}
\begin{figure}
\begin{center}
\includegraphics[width=7cm]{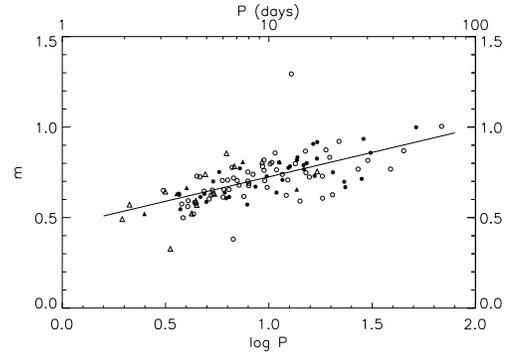}
\end{center}
\vspace*{-5mm}
\caption{Plot of the $m$ parameter vs. $\log P$. The meaning of the symbols 
is the same as in Figs.~\ref{fig_p-a} and~\ref{fig_ar}.}
\label{fig_slope}
\end{figure}

The definition of the $m$ parameter is visualized in Fig.~\ref{fig_slope_def}
showing normalized amplitudes of four Cepheids: two short-period and two 
long-period ones. Each pair consists of a solitary Cepheid (V~Lac and TT~Aql, 
respectively), while the other Cepheid belongs to a binary system 
(SU~Cyg and SS~CMa, respectively) and their pulsation period is pairwise 
similar. This figure clearly shows that the presence of a blue companion 
reduces the slope of the fitted line. The $m$ parameter is also period 
dependent, which is obvious from both Figs.~\ref{fig_slope_def} and~\ref{fig_slope}.

The period dependence is related to the temperature dependence of the 
instability region for various luminosities. The influence of temperature 
on the value of $m$ is supported by the uniform behaviour of 
Cepheids pulsating in different (either fundamental or first overtone) 
modes in the $m$ versus $\log P$ graph (Fig.~\ref{fig_slope}). 
The linear least squares fit to the 127 data points in this graph 
(SU~Cru has been excluded) provides the relation

\begin{equation}
m = (0.282 \pm 0.027) \times \log P + 0.440 \pm 0.027
\end{equation}

Practically the same equation is obtained by taking into account 
only solitary Cepheids. This infers that, in addition to blue companions identified efficiently by UV-spectroscopy (Evans \cite{E92}), 
red stars also form binaries together with Cepheid variables. 
Red secondaries are, however, not as luminous as high temperature 
companions because of the well-known distribution of stars 
in the H-R diagram. This means that most of these red stars
may be line-of-sight optical companions unrelated to the given Cepheid.

The points showing largest deviations from the ridge-line fit 
in Fig.~\ref{fig_slope} correspond mostly to known binaries 
(e.g., SU~Cyg, V1334~Cyg, LS~Pup). 
There are, however, outlying points representing Cepheids classified 
as solitary stars in Table~1. Based on their position in Fig.~\ref{fig_slope}, 
\object{VY~Cyg}, \object{VZ~Pup}, \object{RY~Vel}, and \object{SW~Vel} 
may have a blue companion, while for \object{SZ~Aql}, \object{X~Cyg}, 
and \object{KQ~Sco}, a red companion is inferred.

\subsubsection{The $k$ parameter}
\label{ssect_cawd}

The advantage of the $m$ parameter is its simple definition. 
Omission of the $I$ amplitude, an important piece of information, 
is, however, a drawback. When including the $I$ band amplitude 
in addition to the four other ($U$, $B$, $V$, and $R$) amplitudes, 
the dependence of photometric amplitude on the wave number, $1/{\lambda}$, 
is no longer linear. The graph of photometric amplitudes, 
$A_{\lambda}$, as a function of the effective wavelength, 
$\lambda$, can be approximated well by a function of the type
\begin{equation}
A_{\lambda} = m_k\times(1/{\lambda})^k + {\rm const.}
\label{eq-cawd}
\end{equation}

\begin{figure}
\begin{center}
\includegraphics[width=7cm]{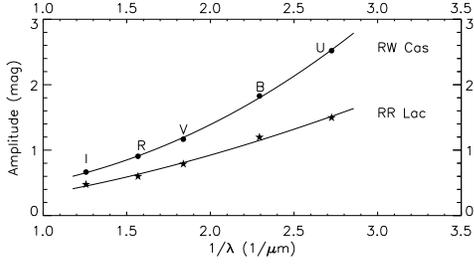}
\end{center}
\vspace*{-3mm}
\caption{Dependence of photometric amplitudes on the observational wavelength 
that defines the $k$ parameter (see text).}
\label{fig_cawd_def}
\end{figure}

Here the exponent, $k$, is a useful numerical parameter characterizing the decrease in photometric amplitudes with increasing wavelength. Examples visualizing the effectiveness of fitting this function to the observed 
amplitudes are shown in Fig.~\ref{fig_cawd_def}, where two solitary Cepheids 
(the short period RR~Lac and the long period RW~Cas) serve as templates. 
It is obvious that the $m$ and $k$ parameters dependent on each other.

\begin{figure}
\begin{center}
\includegraphics[width=7cm]{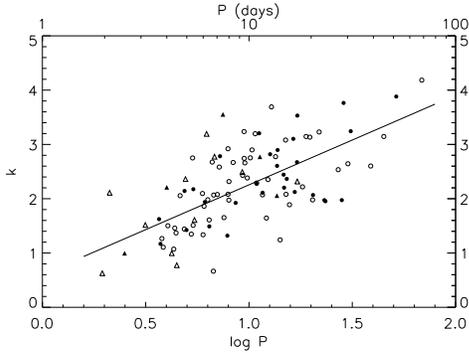}
\end{center}
\vspace*{-5mm}
\caption{Plot of $k$ vs. $\log P$. The meaning of the symbols is the same as 
in Figs.~\ref{fig_p-a} and~\ref{fig_ar}.}
\label{fig_cawd}
\end{figure}

\begin{figure}[h]
\begin{center}
\includegraphics[width=7cm]{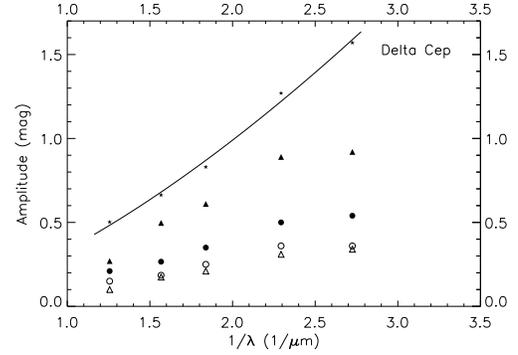}
\end{center}
\vspace*{-3mm}
\caption{Strangely low $U$ amplitude of representative 
double-mode Cepheids. Triangles denote amplitudes of TU~Cas, 
while circles refer to amplitudes of EW~Sct. Filled and open symbols 
represent values for the fundamental mode and first overtone, respectively. 
Amplitudes of $\delta$~Cephei are plotted for comparison purposes.}
\label{fig_cawd-beat}
\end{figure}

The dependence of the $k$ parameter on the decimal logarithm 
of the pulsation period is shown in Fig.~\ref{fig_cawd}. 
The linear least squares fit to the 114 data points in
this graph corresponds to the relation:
\begin{equation}
k = (1.650 \pm 0.173) \times \log P + 0.606 \pm 0.172
\end{equation}
\label{fit-cawd}
We note that $k$ values of the beat Cepheids TU~Cas and
EW~Sct (listed in Table~\ref{beatk}) are much lower than the 
corresponding values determined for single-mode Cepheids pulsating 
with similar period. As can be seen in Fig.~\ref{fig_cawd-beat}, 
this anomaly is caused mainly by the relatively low $U$ amplitude. 
This might indicate a blue companion to both TU~Cas and EW~Sct 
(without any additional evidence of a companion) 
or a strange partition in the pulsational energy 
between the main modes and coupling terms. This latter option 
has to be studied by hydrodynamic modelling of the pulsating atmosphere.

\begin{table}[h]
\caption{Amplitudes of double-mode Cepheids TU~Cas and EW~Sct}
\begin{tabular}{l@{\hskip3mm}c@{\hskip3mm}c@{\hskip2mm}ccc@{\hskip3mm}c@{\hskip2mm}c}
\hline
\noalign{\smallskip}
Type of & \multicolumn{3}{c}{TU Cassiopeiae} & & \multicolumn{3}{c}{EW Scuti} \\
\noalign{\smallskip}
\cline{2-4} \cline{6-8}  
\noalign{\smallskip}
Ampl. & $A_{\rm F}$ & $A_{\rm 1OT}$ & $A_{\rm 1OT}/A_{\rm F}$ & & $A_{\rm F}$ & $A_{\rm 1OT}$ & $A_{\rm 1OT}/A_{\rm F}$\\
\hline
\noalign{\smallskip}
$A_U$ & 0.92 & 0.34 & 0.37 && 0.54 & 0.36 & 0.67\\
$A_B$ & 0.89 & 0.31 & 0.35 && 0.50 & 0.36 & 0.72\\
$A_V$ & 0.61 & 0.21 & 0.34 && 0.35 & 0.25 & 0.71\\
$A_R$ & 0.43 & 0.15 & 0.35 && 0.23 & 0.16 & 0.70\\
$A_I$ & 0.27 & 0.10 & 0.37 && 0.21 & 0.15 & 0.71\\
$A_{V_{\rm RAD}}$ & 29.1 & 13.3 & 0.46 && 13.5 & 13.3 & 0.99\\
\noalign{\smallskip}
\hline
\end{tabular}
\label{beatamp}
\end{table}

\begin{table}[h]
\caption{$k$ values for each mode of TU~Cas and EW~Sct}
\begin{tabular}{l@{\hskip3mm}c@{\hskip3mm}cc}
\hline
\noalign{\smallskip}
Mode & TU Cassiopeiae & & EW Scuti \\
\noalign{\smallskip}
\hline
\noalign{\smallskip}
F & $k=0.038 \pm 0.100$ && $k=0.014 \pm 0.084$  \\
1OT & $k=0.020 \pm 0.091$ && $k=0.013 \pm 0.112$ \\
\noalign{\smallskip}
\hline
\end{tabular}
\label{beatk}
\end{table}

The $k$ parameter is also a good indicator of duplicity
if the temperature of the companion differs from that of the Cepheid.
Most points showing largest deviations from the ridge-line fit 
in Fig.~\ref{fig_cawd} correspond to known binaries (e.g., SU~Cyg, LS~Pup). 
Other outlying points from the fitted line implicitly infer the presence 
of a companion. The Cepheids classified as solitary stars are 
VY~Cyg, VZ~Pup, RY~Vel, SW~Vel (blue companion is inferred), 
SZ~Aql, V1344~Aql, KQ~Sco, and DR~Vel (red companion is inferred). 
Both the $m$ and $k$ parameters of CR~Cep and V495~Cyg are indicative of 
a blue companion, which is consistent with their classification 
as binaries (see Sect.~\ref{ssect_ar}).

\section{Conclusion}
\label{sect_concl}

The dependence of amplitudes on period and interdependences of various 
amplitudes have been investigated based on a homogeneous data set of 
observed pulsational amplitudes of 369 Galactic Cepheids. 
Period--amplitude plots were compiled for each of the $U$, $B$, $V$, $R$, 
and $I$ photometric amplitudes and for the amplitude of the pulsational 
radial velocity variations. An upper envelope was determined for each 
period-amplitude plot. These envelopes enable us to study regularities 
in the behaviour of the pulsation amplitudes using the subsample of 
solitary Cepheids (over 200 stars, 60\% of the entire sample) since 
companions falsify observable amplitudes.

The wide range of observable amplitudes at any pulsation period seen 
in Fig.~\ref{fig_p-a} is only partly caused by companions. 
As pointed out by Sandage et~al. (\cite{Sandetal04}) and Turner et~al.
(\cite{Tetal06a}), photometric amplitudes depend on the position of the
Cepheid in the instability strip, and this effect causes a spread in
amplitudes at a given pulsation period. Nevertheless, companions
also contribute to the amplitude range. Cepheids with luminous 
blue companions (e.g., KN~Cen, AX~Cir, SY~Nor, SV~Per) have much 
smaller photometric amplitudes than the value defined by the 
upper envelope at the given pulsation period. Relatively low 
amplitudes, however, do not necessarily infer binarity.

Cepheids pulsating in the fundamental mode tend to oscillate
with an amplitude that does not differ too much from the possible 
maximum value. However, no such preference is apparent for Cepheids 
oscillating in the first overtone.

Based on the observed amplitudes, three numerical parameters have been 
introduced. The period dependence of these parameters, referred to as 
$q$, $m$, and $k$ (see Sect.~\ref{ssect_ar}, \ref{ssect_slope}, 
and \ref{ssect_cawd}, respectively) has been investigated.

Although pulsation theory predicts that $q$ is a good indicator of the 
mode of pulsation, $q$ values determined from the observed amplitudes 
are not reliable indicators of the pulsation mode. Exceptionally large $q$ 
values  (for a given mode) are, however, indicative of duplicity of the given 
Cepheid. The much smaller than predicted separation between the average
$q$ values of fundamental mode and first overtone Cepheids may be caused,
at least in part, by assigning the incorrect pulsation mode to a number of 
Cepheids. Fundamental mode pulsators can be present among s-Cepheids 
(e.g., \object{SZ~Cas}, \object{AZ~Cen}, \object{BB~Cen}, \object{V1726~Cyg}, 
\object{Y~Oph}) that are usually claimed to be pulsating in the first overtone. 
This view is shared by Bersier \& Burki (\cite{BB96}). Turner et~al. 
(\cite{Tetal06b}) also doubt that \object{V1726~Cyg} pulsates 
in the first overtone: its carefully determined luminosity 
is consistent with that of a fundamental mode pulsator. The existence of 
fundamental mode Cepheids pulsating with small amplitude is also predicted 
by theoretical computations (Szab\'o et~al. \cite{Szetal07}).

The $m$ and $k$ parameters characterizing the wavelength dependence of 
the observed amplitudes can be used to identify Cepheids with either 
blue or red companions. Both $m$ and $k$ parameters linearly depend 
on $\log P$, therefore their true values must be related to the 
typical values valid for the given pulsation period in duplicity studies.

Quite a few Cepheids classified as solitary stars may have a companion. 
These Cepheids and the indication of their duplicity by various amplitude 
parameters are listed in Table~\ref{binind}. The duplicity of VW~Cas, 
CR~Cep, and V495~Cyg cannot be doubted based on ample indications. 
Therefore, a binarity flag 1 has been assigned to these three Cepheids 
in Table~1. We note, however, that physical relationship between a Cepheid 
and its companion (i.e., binarity) can be taken as certain 
if the orbital motion can be identified from radial velocity or astrometric
data, or from a strictly periodic light-time effect in the $O-C$ diagram.
New spectroscopic studies are recommended for all Cepheids mentioned 
in Table~\ref{binind}. Other candidate binaries involving a bright 
Cepheid component (e.g., V950~Sco, LR~TrA) not appearing in 
Table~\ref{binind} also deserve spectroscopic observations.

\begin{table}[]
\caption{New indication of binarity from the amplitude behaviour. 
A + sign means a value indicating a companion. Letter `r' denotes a 
red companion, otherwise blue companions are inferred (except in the column $q$). 
`na' stands for non-applicable because of missing amplitude(s).}
\begin{tabular}{lcccc}
\hline
\noalign{\smallskip}
Cepheid & $q$ & $A_B/A_V$ & $m$ & $k$ \\
\noalign{\smallskip}
\hline
\noalign{\smallskip}
SZ Aql    & & & +r & +r \\
V1344 Aql & & & & + \\
CY Car    & & + & na & na \\
FM Car    & na & + & na & na \\[1ex]
UZ Cas    & + & + & na & na \\
VW Cas    & + & & + & + \\
BP Cas    & + & + & na & na \\
CT Cas    & + & & na & na \\[1ex]
AY Cen    & & + & na & na \\
CR Cep    & + & & + & + \\
X Cyg     & & & +r & \\
VY Cyg    & & & + & + \\[1ex]
GI Cyg    & na & + & na & na \\
V459 Cyg  & & + & na & na \\
V495 Cyg  & + & + & + & + \\
V924 Cyg  & & + & na & na \\[1ex]
UY Per    & & + & na & na \\
VZ Pup    & & & + & + \\
V773 Sgr  & & + & & na \\
KQ Sco    & & & +r & +r \\[1ex]
RY Vel    & & + & + & + \\
SW Vel    & & & + & + \\
DR Vel    & & & & +r \\
\noalign{\smallskip}
\hline
\end{tabular}
\label{binind}
\end{table}

A dichotomy was found separating short- and long-period Cepheids
(Figs.~~\ref{fig_dichotomy} and~\ref{amp_amp}). This dichotomy was 
already found for the $A_B/A_V$ amplitude ratio by van Genderen 
(\cite{vG74}). We have identified similar dichotomies in other 
amplitude ratios, including $A_{V_{\rm RAD}}/A_V$. This dichotomy was 
also found in the colour-magnitude diagrams of Cepheids in our Galaxy 
and the LMC (Sandage et~al. \cite{Sandetal04}), and in the $P$-$L$ and 
period-colour relationships for Cepheids in the LMC (Ngeow et~al. \cite{Netal05}). A break at $\log P \approx 1$ also appears in the 
$A_V$ versus $\langle V-I \rangle_0$ diagram of Cepheids in our Galaxy 
and both Magellanic Clouds (Kanbur \& Ngeow \cite{KN04}, \cite{KN06}).

We determined the limiting period separating short- and
long-period Cepheids to be 10\fd47 (i.e., $\log P = 1.02$).
This period limit is definitely longer than the fundamental period
corresponding to the resonance centre for Galactic Cepheids as is
obvious from Fig.~\ref{fig_p-a}, so the dichotomy and resonance in
radial stellar oscillations are unrelated phenomena.

The currently used value of exactly 10 days for the break in the 
$P$-$L$ relationship can be accepted only in terms of an anthropomorphic viewpoint. The realistic value of 10\fd47 will certainly provide more
reliable $P$-$L$ relations.

Finally, we note that when deriving photometric amplitudes, we observed 
that phase curves of some Galactic s-Cepheids plotted using the 
correct pulsation period exhibit wider scatter than expected from 
measurement errors. We suspect that a non-radial mode is weakly excited 
in these stars, similarly to some Cepheids in the LMC showing such 
phenomenon (Moskalik \& Ko{\l}aczkowski \cite{MK09}).

\begin{acknowledgements}
Financial support from the OTKA grant T046207 is gratefully acknowledged. 
This research was supported by the European Space Agency (ESA) and the
Hungarian Space Office via the PECS programme (contract No.\,98090).
This research has made use of the {\em SIMBAD} database, operated at CDS, 
Strasbourg, France. Aliz Derekas kindly made at our disposal her unpublished 
radial velocity data on a number of southern Cepheids. 
The authors are indebted to Drs.~Szil\'ard Csizmadia, Martin Groenewegen, 
M\'aria Kun, and Chow-Choong Ngeow for their constructive remarks, 
to David Westley Miller for correcting the English text,
and to the referee, Dr. David G.~Turner, whose enlightening and 
critical report helped improve the presentation of the results.
\end{acknowledgements}


\end{document}